# Alleviating Interview Stress: Effects of Robot Persona and Reactive Backchanneling on High-Pressure Mock Interviews

Haochen Yang, Zhuoran Gao, Geqing Tao, Yusi Dong and Pei-Luen Patrick Rau*

*Abstract*— **High-pressure job interviews induce anxiety and impair performance. Existing mock training suffers from a trade-off among cost, standardization, and physical presence. This study investigated how a robot interviewer's verbal persona (supportive vs. challenging) and nonverbal backchanneling (active vs. negative) affect users' psychological states and facial expressions in a simulated interview. In a 2×2 between-subjects experiment (N=28), participants interacted with a NAO robot. Subjective measures (state anxiety, self-efficacy, social presence) and objective facial emotion recognition were collected. Results showed that a supportive persona reduced task load, increased perceived robot warmth and mind perception, and decreased perceived surveillance and robot discomfort. Active backchanneling increased non-neutral ($p = .021$) and anxiety-tension facial affect ($p = .048$). A significant interaction effect between persona and backchanneling was found on social presence ($p = .045$). The subjective-objective consistency was 67.9%, indicating complementary measurement modalities. The findings provide preliminary support for the social buffering hypothesis in human-robot interaction.**

## I. Introduction

Job interview is a high-stakes social evaluative context. Interview anxiety impairs fluency, self-presentation, and performance [1], creating a demand for low-cost, repeatable mock interview training. Existing training falls into human-based simulation (costly, non-standardized, stressful) and screen/voice systems (lacking physical presence and nonverbal dynamics [2]). Embodied robots offer a balanced solution: socially assistive robots like NAO have shown positive effects in interaction and instruction [3] with good controllability.

Human-robot interaction (HRI) research has shown that a robot's persona affects users' affect and trust [4], [5]. In high-pressure interviews, nonverbal backchanneling (nodding, gaze) also plays a critical role – in human-human interaction, nonverbal responsiveness enhances rapport and social comfort; in HRI, gaze and timely feedback increase social presence [6]. Yet little research has examined whether backchanneling can serve as a social buffer to mitigate task load and emotional tension in challenging interviews.

This study draws on a multi-theory framework. The CASA (Computers Are Social Actors) paradigm suggests that humans unconsciously treat media as social actors [7], providing a basis for cross-modal integration. According to the buffering hypothesis [8] and self-efficacy theory [9], a supportive verbal persona is expected to reduce task load, increase perceived robot warmth and mind perception, and decrease perceived surveillance and robot discomfort, compared to a challenging persona (H1). Based on the social buffering perspective, active backchanneling is hypothesized to increase emotional engagement, reflected in greater non-neutral facial affect change relative to negative backchanneling (H2). Furthermore, according to Expectancy Violations Theory (EVT) [10] and the Uncanny Valley of Mind [11], the congruence between verbal persona and nonverbal backchanneling may critically shape user experience. Specifically, congruent combinations (e.g., supportive persona with active backchanneling, or challenging persona with negative backchanneling) are expected to enhance social presence, whereas incongruent combinations may impair it. The effect of backchanneling on social presence is moderated by persona: active backchanneling will increase social presence under a supportive persona but decrease it under a challenging persona (H3).

Using a 2×2 between-subjects experiment, this pilot study investigates how verbal persona (supportive vs. challenging) and backchanneling type (active vs. negative) jointly influence subjective states and objective emotional expression, providing preliminary empirical support for the social buffering hypothesis in HRI.

## II. Method

### A. Participants

A total of 28 students (14 male, 14 female; age 19–22, M = 20.64, SD = 0.83) participated. All participants were native Chinese speakers with normal vision and hearing. They gave informed consent and were randomly assigned to four experimental conditions, with seven participants per condition.

### B. Experimental Design

This study employed a 2 (persona: supportive vs. challenging) × 2 (backchanneling: active vs. negative) between-subjects design. For the supportive persona, the agent used encouraging and warm phrasing (e.g., "Your approach is very clear, which is highly commendable!"). For the challenging persona, it used critical and pressuring language (e.g., "What do you mean? Don't be ambiguous."). Active backchanneling consisted of timely nodding, gaze, and other listening behaviors triggered by speech pauses; negative backchanneling involved only a basic idle posture. The four

* Research supported by Department of Industrial Engineering, Tsinghua University.

Haochen Yang, Zhuoran Gao, Geqing Tao, Yusi Dong and Pei-Luen Patrick Rau are with Department of Industrial Engineering, Tsinghua University, Beijing 100084, People's Republic of China (corresponding author to provide phone: 010-6277-6664; e-mail: rpl@tsinghua.edu.cn).

conditions are illustrated in Figure 1 .

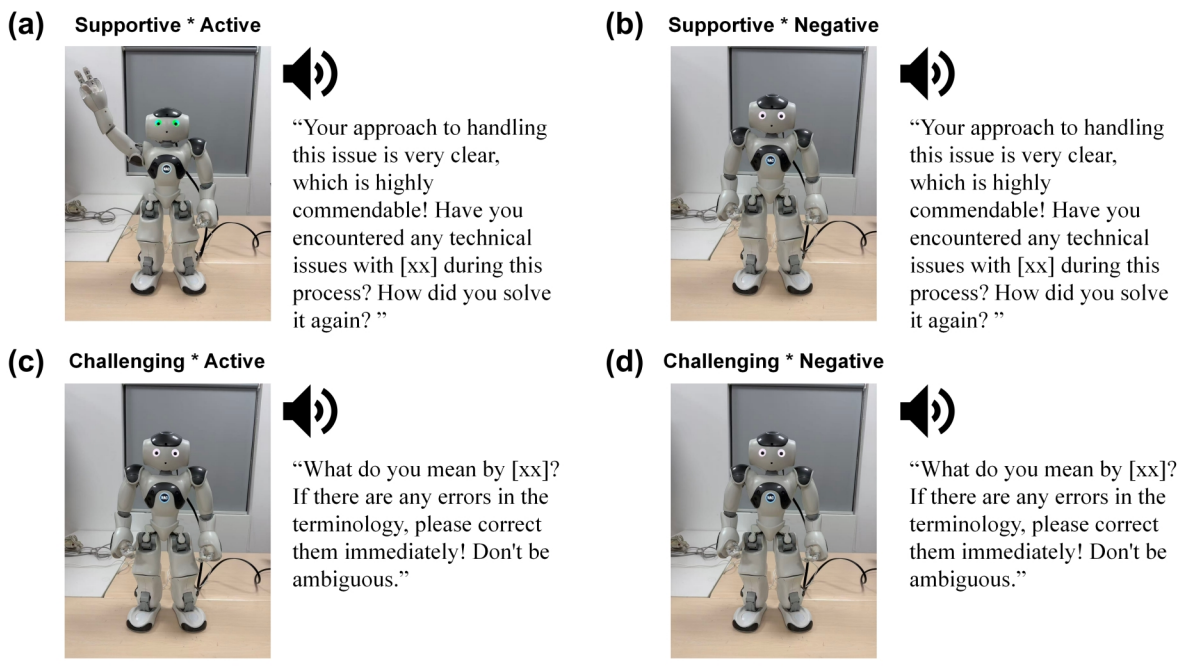

Figure 1. Illustration of the four experimental conditions. (a) Supportive persona with active backchanneling; (b) Supportive persona with negative backchanneling; (c) Challenging persona with active backchanneling; (d) Challenging persona with negative backchanneling. Example actions and utterances from the robot are shown for each condition.

### C. *Apparatus and Technical Architecture*

The experimental system comprised several devices. A NAO V6 robot served as the physical interviewer. A smartphone with the Doubao app (pre-configured with supportive or challenging persona) played verbal feedback via a Bluetooth speaker behind the NAO. A DJI Action 4 camera recorded facial expressions. A separate computer displayed instructions/questionnaires; another computer ran a Python console for the experimenter to manually trigger NAO's nonverbal movements per condition. Thus, the system adopted a decoupled operation: verbal persona was controlled by the Doubao agent, whereas backchanneling was implemented via Wizard-of-Oz control of the robot's body (see Figure 2).

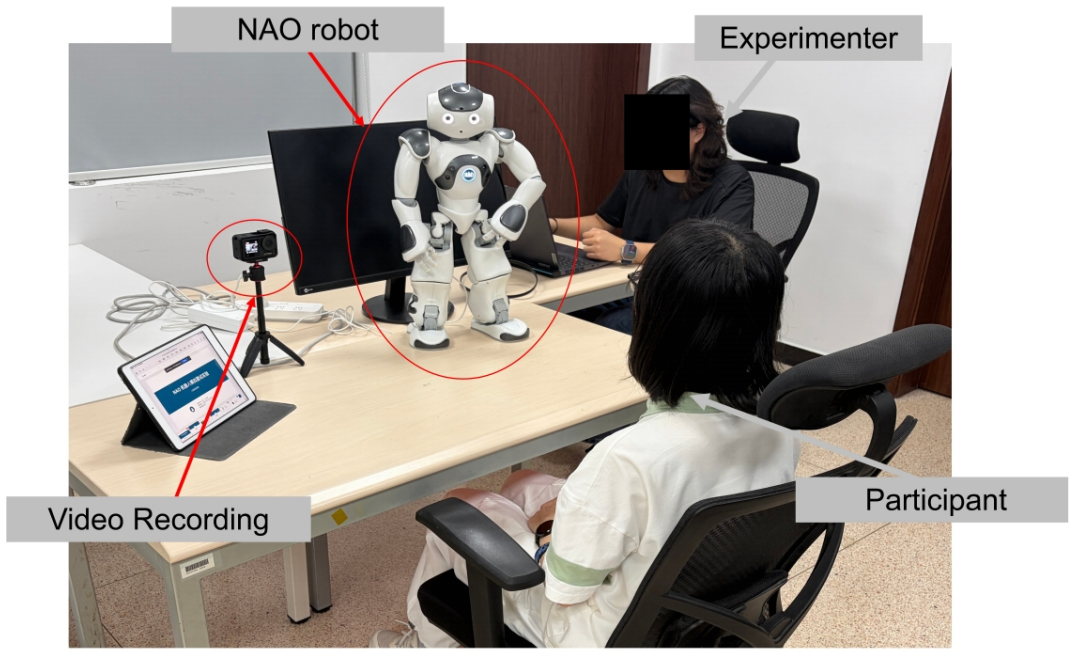

Figure 2. Experimental setup and arrangement of devices/personnel

### D. *Procedure*

The experimental procedure consisted of three phases. In the first phase (pre-test and baseline collection), participants completed questionnaires on demographics and trait anxiety (STAI-T-6) [12], and engaged in a standardized 2-minute small talk (e.g., "Please briefly introduce your hometown") to capture baseline facial emotion expression. In the second phase (formal interview), participants completed a 6–8 minute mock interview with the NAO robot on the topic of a summer internship position. The robot's verbal persona and backchanneling type were controlled according to the participant's assigned condition. Video recording was conducted simultaneously throughout this phase. In the third phase (post-test), participants completed the post-questionnaire (all scales summarized in TABLE I) along with manipulation check items.

### E. *Dependent Variables and Covariates*

All subjective dependent variables in this study were measured using 7-point Likert scales. Specifically, state anxiety was assessed with the Chinese short version of the STAI-S-6 [12]; interview self-efficacy was measured with a custom three-item scale [10], [13]; robot social presence was measured with the eight-item short version of Chen et al. [14]; robot social attributes were measured with the RoSAS-SF [15], [16]; expectancy violation valence and mind perception [11], [17] were custom or adapted scales; cognitive load was measured with the NASA-TLX [18]; and sense of being watched was measured with a custom single item [19]. Manipulation checks included two items: perceived persona and perceived backchanneling.

The objective dependent variable was derived from video analysis. Facial emotion expression was recognized using the DeepFace engine for seven basic emotions (anger, disgust, fear, happiness, sadness, surprise, neutral). Composite indicators such as Negative Affect (sad + angry + fear + disgust) and Positive Affect (happy + surprise) were then computed.

Covariates in this study included baseline facial emotion expression collected during the small-talk phase, and trait anxiety (STAI-T-6 [12]). These covariates were used in subsequent covariance analyses to adjust for individual differences.

TABLE I. DEPENDENT VARIABLES AND COVARIATES

| Type | Variable | Measure / Dimension | α |
|---|---|---|---|
| Subjective DVs | State Anxiety | STAI-S-6 | 0.956 |
| | Interview Self-Efficacy | Custom scale (3 items) | 0.886 |
| | Robot Social Presence | Custom scale (8 items) | 0.855 |
| | Expectancy Violation Valence | Custom scale (4 items) | 0.847 |
| | Mind Perception - Experience | Custom scale (3 items) | 0.733 |
| | Mind Perception - Agency | Custom scale (2 items) | 0.586 |
| | Cognitive Load | NASA-TLX | 0.794 |
| | Sense of Being Watched | Custom scale (1 items) | — |
| | Robot Social Attributes - Warmth | RoSAS-SF | 0.808 |
| | Robot Social Attributes - Competence | RoSAS-SF | 0.834 |
| | Robot Social Attributes - Discomfort | RoSAS-SF | 0.855 |
| Objective DV | Facial Emotion Expression | DeepFace recognition (7 emotions) | — |
| Covariates | Trait Anxiety | STAI-T-6 | 0.693 |
| | Baseline Facial Emotion | Small- talk phase video analysis | — |

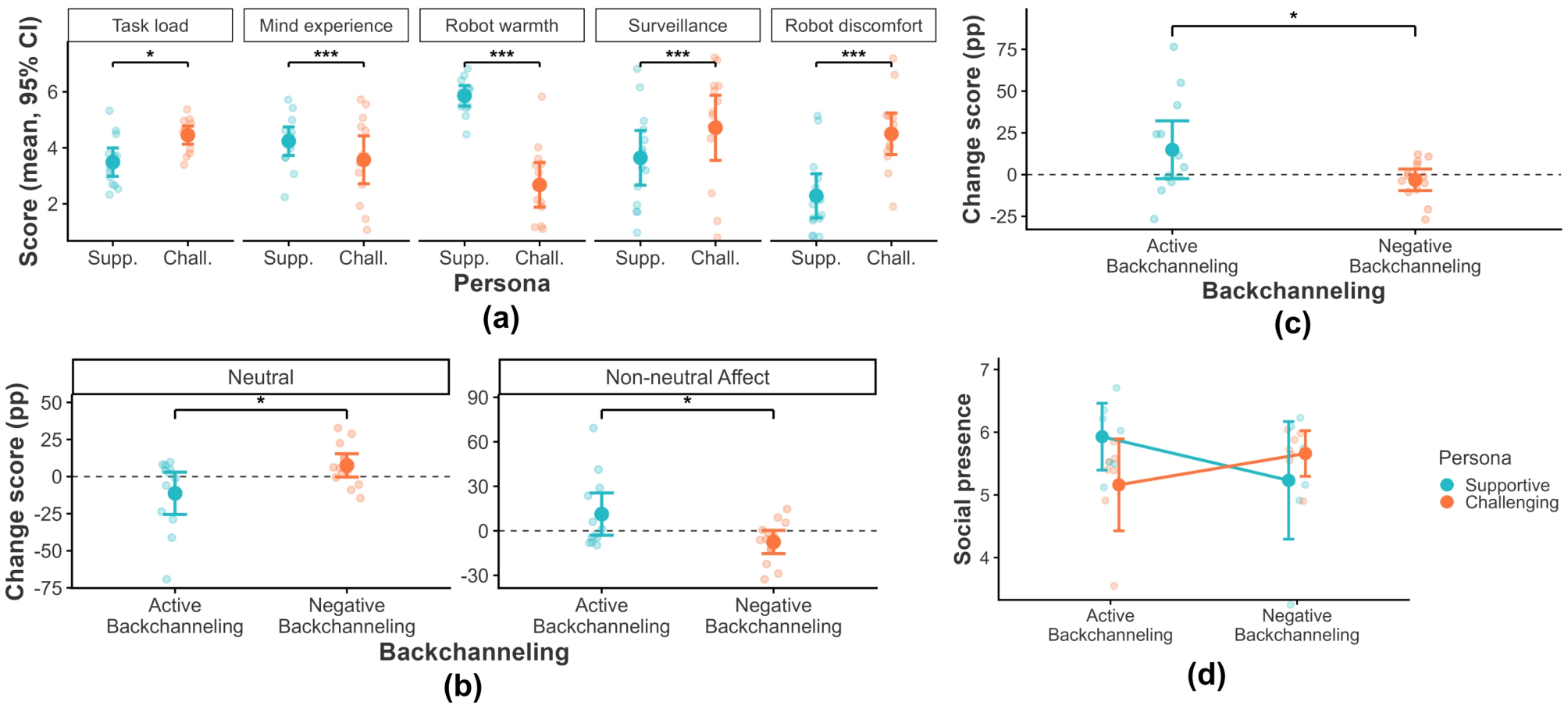

Figure 3. Significant main and interaction effects of persona and backchanneling. (a) Main effects of persona on the five subjective measures. (b) Main effect of backchanneling type on facial affect change (non-neutral vs. neutral). (c) Main effect of backchanneling type on anxiety-tension facial affect change (fear + sad). (d) Interaction effect of persona and backchanneling type on perceived social presence.

## III. Result

### A. Manipulation Checks

Independent-samples t-tests confirmed successful manipulations. For persona, participants rated the challenging robot as significantly more challenging (M = 6.29, SD = 0.73) than the supportive robot (M = 1.43, SD = 0.65), $t(26) = -22.76$, $p < .001$, Cohen's $d = 8.60$. For backchanneling, the active condition was perceived as showing more listening behaviors (M = 6.07, SD = 1.00) than the negative condition (M = 5.00, SD = 1.30), $t(26) = 2.26$, $p = .035$, Cohen's $d = 0.85$.

### B. H1: Effects of Persona

ANCOVAs with baseline measures as covariates were conducted on the subjective dependent variables. Significant main effects of persona were observed for five measures.

Participants interacting with the supportive persona reported lower task load (M = 3.55, SD = 0.92) than those with the challenging persona (M = 4.32, SD = 0.65), $p = .017$. Higher mind perception (Experience) was found for the supportive persona (M = 5.26, SD = 0.81) compared to the challenging persona (M = 3.67, SD = 1.31), $p < .001$. Similarly, robot warmth was rated higher for the supportive persona (M = 5.39, SD = 0.81) than for the challenging persona (M = 3.07, SD = 1.99), $p < .001$. Conversely, participants in the challenging persona condition reported higher perceived surveillance (M = 4.93, SD = 1.36) than those in the supportive persona condition (M = 3.00, SD = 1.14), $p < .001$, and greater robot discomfort (M = 4.64, SD = 1.08) relative to the supportive persona (M = 2.32, SD = 1.57), $p < .001$ (see Figure 3 (a)). No significant main effect of persona was found for state anxiety, self-efficacy, or other subjective measures.

### C. H2: Effects of Backchanneling

To examine H2, facial-expression change scores (interview-stage affect minus baseline affect) were computed. A change-score ANOVA revealed a significant main effect of backchanneling type on non-neutral facial affect change. Participants in the active backchanneling condition showed a larger increase in non-neutral affect (M = 69.24%, SD = 23.66%) than those in the negative condition (M = 56.24%, SD = 31.82%), $p = .021$ (see Figure 3 (b)). An exploratory analysis using the anxiety-tension composite (fear + sad) also revealed a significant main effect of backchanneling type, with active backchanneling leading to greater increases (M = 57.04%, SD = 31.33%) than negative backchanneling (M = 51.63%, SD = 31.14%), $p = .048$ (see Figure 3 (c)). No significant main effect of backchanneling was found for any subjective measures.

### D. H3: Interaction Between Persona and Backchanneling

A two-way ANCOVA on perceived social presence yielded a significant interaction between persona and backchanneling type, $p = .045$ (see Figure 3 (d)). Under the supportive persona condition, active backchanneling was associated with higher social presence (M = 6.07, SD = 0.60) than negative backchanneling (M = 5.09, SD = 1.22). Under the challenging persona condition, the pattern reversed: active backchanneling led to lower social presence (M = 5.02, SD = 1.16) compared to negative backchanneling (M = 5.61, SD = 0.45).

TABLE II. Persona and Backchanneling Effects

| Hypothesis | Outcome | $F$ [a] | $p$ | $\eta_p^2$ |
|---|---|---|---|---|
| H1 | Task load | 6.60 | .017 | 0.480 |
| H1 | Mind perception - Experience | 16.62 | < .001 | 0.683 |
| H1 | Robot warmth | 19.54 | < .001 | 0.675 |
| H1 | Perceived surveillance | 16.14 | < .001 | 0.934 |
| H1 | Robot discomfort | 20.27 | < .001 | 0.922 |
| H2 | Non-neutral facial affect change | 6.19 | .021 | 0.212 |
| H2 | Anxiety-tension facial affect change | 4.38 | .048 | 0.160 |
| H3 | Social presence | 4.49 | .045 | 0.296 |

a. All F-tests have df = (1, 23).

## E. Subjective-Objective Consistency

Correlation analyses revealed weak, non-significant relationships between subjective state anxiety and objective facial expression changes. Fear expression change correlated weakly with subjective state anxiety ($r$ = 0.255, $p$ = .208), while negative affect change showed almost no correlation ($r$ = 0.005, $p$ = .982). Happy expression change was not correlated with interview self-efficacy ($r$ = −0.057, $p$ = .784).

The consistency rate between subjective and objective measures was 67.9%, indicating that 32.1% of participants showed mismatched subjective-objective anxiety responses (i.e., one measure increased while the other decreased).

# IV. Conclusion

This study examined how a robot's verbal persona and nonverbal backchanneling affect users in a high-pressure mock interview.

First, H1 was supported. Compared to a challenging persona, the supportive persona reduced task load, increased warmth and mind perception, and decreased surveillance and discomfort. Although high baseline stress likely limited subjective anxiety reduction within this brief interaction, the robot successfully buffered cognitive and interpersonal dimensions. Second, H2 was supported. Active backchanneling (nodding, gaze) increased non-neutral and anxiety-tension facial affect relative to negative backchanneling, indicating that it enhances facial expressivity rather than simply reducing anxiety. Third, H3 was supported. Active backchanneling improved social presence under the supportive persona but backfired under the challenging persona, a pattern that aligns with Expectancy Violations Theory and the Uncanny Valley of Mind: congruent verbal-nonverbal signals amplify positive experiences, whereas mismatched signals can worsen perceptions. Additionally, the moderate consistency (67.9%) suggests that subjective and objective measures capture distinct aspects of experience, supporting multi-modal assessment.

As a pilot study, the small sample size limits statistical power; moreover, the single-culture (Chinese) sample and the Wizard-of-Oz control of backchanneling restrict generalizability and autonomy. Future work will validate these findings with larger, cross-cultural samples and fully autonomous, real-time backchanneling systems. Together, these findings provide initial support for the social buffering hypothesis in HRI and offer practical insights for designing robot interview coaches that balance verbal pressure with supportive nonverbal feedback.

# References

[1] J. McCarthy and R. Goffin, "Measuring Job Interview Anxiety: Beyond Weak Knees and Sweaty Palms," *Pers. Psychol.*, vol. 57, no. 3, pp. 607–637, Sep. 2004, doi: 10.1111/j.1744-6570.2004.00002.x.

[2] W. A. Bainbridge, J. Hart, E. S. Kim, and B. Scassellati, "The effect of presence on human-robot interaction," in *RO-MAN 2008 - The 17th IEEE International Symposium on Robot and Human Interactive Communication*, Munich, Germany: IEEE, Aug. 2008, pp. 701–706. doi: 10.1109/ROMAN.2008.4600749.

[3] A. Robaczewski, J. Bouchard, K. Bouchard, and S. Gaboury, "Socially Assistive Robots: The Specific Case of the NAO," *Int. J. Soc. Robot.*, vol. 13, no. 4, pp. 795–831, Jul. 2021, doi: 10.1007/s12369-020-00664-7.

[4] J. Goetz, S. Kiesler, and A. Powers, "Matching robot appearance and behavior to tasks to improve human-robot cooperation," in *The 12th IEEE International Workshop on Robot and Human Interactive Communication, 2003. Proceedings. ROMAN 2003.*, Millbrae, CA, USA: IEEE, 2003, pp. 55–60. doi: 10.1109/ROMAN.2003.1251796.

[5] A. Tapus, C. Ţăpuş, and M. J. Matarić, "User—robot personality matching and assistive robot behavior adaptation for post-stroke rehabilitation therapy," *Intell. Serv. Robot.*, vol. 1, no. 2, pp. 169–183, Apr. 2008, doi: 10.1007/s11370-008-0017-4.

[6] N. Anzabi and H. Umemuro, "Effect of Different Listening Behaviors of Social Robots on Perceived Trust in Human-robot Interactions," *Int. J. Soc. Robot.*, vol. 15, no. 6, pp. 931–951, Jun. 2023, doi: 10.1007/s12369-023-01008-x.

[7] B. Reeves and C. Nass, "The Media Equation: How People Treat Computers, Television, and New Media Like Real People and Pla," *Bibliovault OAI Repos. Univ. Chic. Press*, Jan. 1996.

[8] B. L. Wilcox, "Social support, life stress, and psychological adjustment: A test of the buffering hypothesis," *Am. J. Community Psychol.*, vol. 9, no. 4, pp. 371–386, Aug. 1981, doi: 10.1007/BF00918169.

[9] A. Bandura, "Guide for Constructing Self-Efficacy Scales," in *Self-Efficacy Beliefs of Adolescents*, F. Pajares and T. Urdan, Eds., Emerald Publishing Limited, 2006, p. 0. doi: 10.1108/978-1-60752-837-120251015.

[10] J. K. Burgoon *et al.*, "Application of Expectancy Violations Theory to communication with and judgments about embodied agents during a decision-making task," *Int. J. Hum.-Comput. Stud.*, vol. 91, pp. 24–36, Jul. 2016, doi: 10.1016/j.ijhcs.2016.02.002.

[11] H. M. Gray, K. Gray, and D. M. Wegner, "Dimensions of Mind Perception," *Science*, vol. 315, no. 5812, pp. 619–619, Feb. 2007, doi: 10.1126/science.1134475.

[12] Q. Du, H. Liu, C. Yang, X. Chen, and X. Zhang, "The Development of a Short Chinese Version of the State-Trait Anxiety Inventory," *Front. Psychiatry*, vol. 13, p. 854547, May 2022, doi: 10.3389/fpsyt.2022.854547.

[13] J. Barlow, C. Wright, and L. Cullen, "A job-seeking self-efficacy scale for people with physical disabilities: Preliminary development and psychometric testing," *Br. J. Guid. Couns.*, vol. 30, no. 1, pp. 37–53, Feb. 2002, doi: 10.1080/030698880220106500.

[14] N. Chen, X. Liu, Y. Zhai, and X. Hu, "Development and validation of a robot social presence measurement dimension scale," *Sci. Rep.*, vol. 13, no. 1, p. 2911, Feb. 2023, doi: 10.1038/s41598-023-28817-4.

[15] C. M. Carpinella, A. B. Wyman, M. A. Perez, and S. J. Stroessner, "The Robotic Social Attributes Scale (RoSAS): Development and Validation," in *Proceedings of the 2017 ACM/IEEE International Conference on Human-Robot Interaction*, Vienna Austria: ACM, Mar. 2017, pp. 254–262. doi: 10.1145/2909824.3020208.

[16] K. L. Neuenswander, A. Dash, P. D. Koya, L. Lin, G. S. R. Gillespie, and S. J. Stroessner, "Measuring Fundamental Aspects of the Social Perception of Robots: Development and Validation of a Shortened Version of the RoSAS (RoSAS-SF)," *Int. J. Soc. Robot.*, vol. 17, no. 6, pp. 1097–1112, Jun. 2025, doi: 10.1007/s12369-025-01251-4.

[17] I. Cucciniello, S. Sangiovanni, G. Maggi, and S. Rossi, "Mind Perception in HRI: Exploring Users' Attribution of Mental and Emotional States to Robots with Different Behavioural Styles," *Int. J. Soc. Robot.*, vol. 15, no. 5, pp. 867–877, May 2023, doi: 10.1007/s12369-023-00989-z.

[18] S. G. Hart and L. E. Staveland, "Development of NASA-TLX (Task Load Index): Results of Empirical and Theoretical Research," in *Advances in Psychology*, vol. 52, P. A. Hancock and N. Meshkati, Eds., North-Holland, 1988, pp. 139–183. doi: 10.1016/S0166-4115(08)62386-9.

[19] S. Scielzo, S. Fiore, F. Jentsch, and N. Finkelstein, "23. Cognition and Collaboration in Hybrid Human–Robot Teams: Viewing Workload and Performance through the Lens of Multimedia Cognitive Load," *Adv. Hum. Perform. Cogn. Eng. Res.*, vol. 7, pp. 329–342, May 2006, doi: 10.1016/S1479-3601(05)07023-2.